\documentclass[prl,aps,twocolumn,groupedaddress,showpacs,floatfix,preprintnumbers,superscriptaddress]{revtex4-1}
\usepackage{amsmath,amssymb,multirow,bm}
\usepackage[colorlinks,linkcolor=blue,anchorcolor=blue,citecolor=blue,urlcolor=blue,filecolor=black]{hyperref}

\newcommand{\beq}{\begin{equation}}
\newcommand{\eeq}{\end{equation}}
\newcommand{\bea}{\begin{eqnarray}}
\newcommand{\eea}{\end{eqnarray}}
\usepackage{graphicx}
\preprint{}

\begin{document}


\title{Fission of relativistic nuclei with  fragment excitation and reorientation}
\author{Carlos A. Bertulani}
\email{carlos.bertulani@tamuc.edu}\affiliation{Department of Physics and Astromomy, Texas A\&M University-Commerce, Commerce, TX 75429, USA }

\author{Yasemin Kucuk}
\email{ykucuk@akdeniz.edu.tr}
\affiliation{Akdeniz University, Science Faculty, Department of Physics, 07058, Antalya, Turkey}

\author{Radomira Lozeva}
\email{radomira.lozeva@csnsm.in2p3.fr}
\affiliation{Universit\'e Paris-Saclay, CNRS/IN2P3, IJCLab, 91405 Orsay, France}

\date{\today}

\begin{abstract}
Experimental studies of fission  induced in relativistic nuclear collisions show a systematic enhancement of the excitation  energy of the primary fragments  by a factor of $\sim 2$, before their decay by fission and other secondary fragments.  Although it is widely accepted that  by doubling the energies of the single-particle states may yield a better agreement with  fission data, it does not prove fully successful, since it is not able to explain yields for light and intermediate mass fragments.  State-of-the-art calculations are successful to describe the overall shape of the mass distribution of fragments, but fail within a factor of 2-10 for a large number of individual yields. Here, we present a novel approach that provides an account of the additional excitation of primary fragments due to final state interaction with the target. Our method is applied to the $^{238}$U + $^{208}$Pb reaction at 1 GeV/nucleon (and is applicable to other energies), an archetype case of fission studies with relativistic heavy ions, where we find that the large probability of energy absorption through final state  excitation of giant resonances in the fragments  can substantially modify the isotopic distribution of final fragments in a better agreement with data. Finally, we demonstrate that large  angular momentum transfers to the projectile and to the primary fragments via the same mechanism imply the need of more elaborate theoretical methods than the presently existing ones. 

\end{abstract}


\maketitle

{\it Introduction.} Nuclear fission is the outcome of a rearrangement leading to a ``run-away'' dynamics involving large-amplitude nuclear motion. Large energies are involved in fission which can be explained with liquid drop models, but experimental observables such as fragment mass yields and their energies are strongly influenced by shell and pairing effects. Fission studies are usually based on stable or long-lived fissile targets irradiated by neutrons, photons, light-charged particles  \cite{Andreyev_2017,Schmidt_2018}, or multinucleon transfer induced by heavy ions (see Ref \cite{HirosePRL119.222501} and references therein).  However, observables obtained at low energies provide a limited understanding of the influence of  several physical variables as well as the path of the process passing by the saddle point to fission. Additional information  has been obtained with radioactive beams, offering a plethora of new experimental advantages \cite{SCHMIDT2000221}. Radioactive nuclei scattered off stable targets allows one to vary nucleon and isospin content of the fissioning nucleus over wide ranges.  First fission studies with relativistic radioactive beams  (above 100 MeV/nucleon) and in inverse kinematics provided insight into dissipative properties of nuclear matter \cite{RodriguezPRC.94.061601} and allowed to map the transition from asymmetric fission in the actinides to symmetric fission in pre-actinide nuclei \cite{SCHMIDT1994313,Aumann1995,Armbruster1996,BENLLIURE1998458,SCHMIDT2000221}. 

We propose a novel approach to compute isotopic distribution of fission fragments in relativistic heavy ion collisions. Using a combination of systematical reaction models, we explain simultaneously the production of heavy, intermediate and light fragments, as well as the fission yields. We consider fragment excitation and re-orientation after the production of primary fragments and their role on the isotopic distribution of fission products. Our predictions can pave the way to use post-excitation of primary fragments as a new technique to study the difficult subject of fission times of excited nuclei \cite{JACQUET2009155}.

{\it Theoretical framework.} The excitation amplitude ${\cal A}_\alpha \left(z, b\right)$ of relativistic projectiles undergoing fission in-flight is obtained from the coupled-channels equations \cite{BertulaniPRL94.072701}
\bea
i\hbar v {\partial {\cal A}_\alpha (z, b) \over \partial z} &=& \sum_{\alpha'} \left< \alpha \left| {\cal M}_{(E/N)L} \right| \alpha' \right> \nonumber \\
&\times&{\cal A}_{\alpha '}(z,b) e^{- {(E_{\alpha '} -E_\alpha)z/ \hbar v}},\label{rcc}
\eea
where $b$ is the impact parameter, $v$ denotes the projectile and fragment velocities, assumed to be nearly constant, and $z$ is the projectile position along the beam direction. ${\cal M}_{EL}$ is the electromagnetic  (EM) operator for electric dipole (E1) and quadrupole (E2) transitions, and ${\cal M}_{NL}$ is the nuclear transition operator for the multipolarity $L$ connecting states $\alpha$ and $\alpha'$ and satisfying  angular momentum and parity selection rules.  

It suffices to consider  the excitation of the isovector giant dipole (IVGDR), isoscalar giant quadrupole (ISGQR) and isovector giant quadrupole (IVGQR) resonances and the double giant dipole resonance (DGDR), with main decay channels leading to the emission of neutrons and fission fragments.   Later we will discuss the role of rotational states. Inclusion of relativistic effects in the nucleus-nucleus dynamics follows  Ref.  \cite{BertulaniPRL94.072701}. The main contribution to nuclear excitation arises from  the excitation of the ISGQR with a matrix element \cite{SATCHLER1987215}
\begin{eqnarray}
&\left<\alpha |{\cal M}_{N2m} | \alpha ' \right>=-\displaystyle{{\delta_{2}}\over{\sqrt{5}}}\ \left<J_{\alpha '} M_{\alpha '}|Y_{2m}| J_{\alpha } M_{\alpha }\right>\nonumber \\
&\times \ \ Y_{2m}(\hat{\mathbf{r}})\ \displaystyle{dU(r)\over dr} ,\label{VfiN}
\end{eqnarray}
where $\delta_{2}$ is the deformation length, $U(r)$ is the scalar Lorentz-boosted nucleus-nucleus potential \cite{BertulaniPRL94.072701}.  The cross sections for the projectile in the final state $\left|\alpha \right>$ are obtained by integration over impact parameters. 

At large impact parameters only EM excitation occurs, while at ``grazing" impact parameters ($b\sim R_p +R_T$), both EM and nuclear excitation are possible. Below grazing impact parameters, much stronger reactions occur where nucleons can be removed (abrasion) due to binary nucleon-nucleon collisions. In the Glauber model  \cite{HufnerPRC12.1888}, the production cross section of a primary fragment with charge and neutron number $(Z_{F},N_{F})$ for a projectile nucleus with charge and neutron number $(Z_{P},N_{P})$  is 
\begin{eqnarray}
\sigma(Z_{F},N_{F})&=&
\left(
\begin{array}{c}Z_{P} \\ Z_{F}
\end{array}
\right)
\left(
\begin{array}{c}N_{P} \\ N_{F}
\end{array}
\right)
\int d^2 b \left[ 1-P_p(b)\right]^{Z_{P}-Z_{F}} \nonumber \\
&\times& P_p^{Z_{F}}(b)\left[ 1-P_n(b)\right]^{N_{P}-N_{F}}P_n^{N_{F}}(b), \label{sigma}
\end{eqnarray} 
where  the binomial coefficients account for all possible ways that $Z_{F}$ protons can be removed from the $Z_{P}$ initial protons of the projectile. A similar counting is made for the  neutrons. $P_p$ ($P_n$) are the probabilities for the survival of a single proton (neutron) of the projectile and the factors containing $(1-P)$ account for the removal probability of the other protons (neutrons).  The probability $P_p$  is   
\begin{eqnarray}
P_p(b)&=&\int dzd^2s \rho_p^P({\bf s},z) \exp\left[ -\sigma_{pp} Z_T\int d^2s \rho_p^T({\bf b-s},z) \right. \nonumber \\
&-&\left. \sigma_{pn} N_T\int d^2s \rho_n^T({\bf b-s},z) \right],  \label{ppb}
\end{eqnarray} 
where the charge and neutron number of the target is denoted by $(Z_{T},N_{T})$, and $\rho_p$ ($\rho_n$) is the proton (neutron) density of projectile and target,  normalized to unity. $\sigma_{np}$ and $\sigma_{pp}$ are the neutron-proton and proton-proton (without Coulomb) total cross sections, taken from a fit to the experimental data \cite{BertulaniConti10}.  A similar expression  is used for $P_{n}$ with the reversed roles of the neutron and proton quantities.
Neutron and proton single-particle densities are obtained with a deformed Woods-Saxon potential assuming that mean-field rearrangements do not have time to occur until long after the abrasion stage. 

The primary fragment excitation energy  is calculated from the particle-hole energies of the configuration relative to its ground state. The density of excited states $\rho(E_x ,Z_F,A_F)$ results from the counting of all combinations of holes consistent with the charge and mass numbers of the fragment, yielding excitation cross sections ${d\sigma/ d E_x} = \rho(E_x ,Z_F,A_F) \sigma(Z_{F},N_{F})$ \cite{HufnerPRC12.1888,GAIMARD1991709,CarlsonPRC.51.252}.  The de-excitation process (ablation) leading to emission of nucleons, light-charged particles, photons, intermediate-mass fragments (IMFs), and fission products, is obtained using the Ewing-Weisskopf model, incorporated in  the ABLA07 code \cite{aleks2009abla07}.  Separation energies and  emission barriers for charged particles are accounted for using the 2016 atomic mass evaluation \cite{ame2016a} and the Bass potential \cite{bass}, respectively. Fission yields are calculated following the dynamical picture reported in Refs. \cite{JURADO2003186,JURADO200514} which has been benchmarked in several works by comparison with isotopic distributions of fission fragments measured in spallation and fragmentation reactions with relativistic nuclei.

The central idea of the present work is that all relevant cross sections can be reliably calculated with existing reaction formalisms, but with the additional inclusion of post-excitation and decay of the primary fragments missing in previously published models. The reorientation of fragments due to the final state interaction with the Coulomb field of the target and the sudden removal of nucleons also leads to an alignment and modification of the angular momentum of the fragments before scission. Reorientation refers to multiple Coulomb excitation changing magnetic substates and  also changing  total angular momentum  \cite{alder:1956:RMP}.  

{\it Application to the $^{238}${\rm U} $+$ $^{208}${\rm Pb} reaction}. 
The $^{238}$U + $^{208}$Pb reaction at 1 GeV/nucleon yielding fission fragments has been studied experimentally \cite{ENQVIST199947}. Projectile singe-particle states were generated with a deformed Woods-Saxon model \cite{KRUPPA198559} with deformation parameter $\beta=0.29$, radius $R_{0}=6.8$ fm, and diffuseness $a=0.6$ fm.  The potential depths were adjusted to yield the last occupied nucleon orbital with binding energy equal to the nucleon separation energy. The lead target density was taken from electron scattering experiments \cite{DEVRIES1987495}, assumed to be the same for protons and neutrons.
Abrasion probabilities were calculated using Eq. \eqref{ppb} with radial and angular wave functions building up single-particle densities for each state and adding the contributions from all occupied states \cite{CarlsonPRC.51.252}. An average over the orientation of the projectile was also performed. For simplicity, only three orientations have been used; one with the major axis along the beam and the other two perpendicular to it. A total abrasion cross section of 8.06 b was obtained using Eq. \eqref{sigma} and adding up all fragments. This cross section is independent of the second stage (ablation) because it includes all possible decay channels. If deformation is neglected, setting  $\beta=0$, the cross section reduces to 7.89 b. Another reduction of the cross section occurs if instead of densities  from individual contributions of single-particle states in Eq. \eqref{sigma}, one uses $^{238}$U total densities from electron scattering data \cite{DEVRIES1987495}. We obtain an abrasion cross section of 7.65 b, which highlights that details of the shell structure can be responsible for modifications of about 5\% of the abrasion cross sections. 

\begin{figure}
\begin{center}
\includegraphics[scale=0.38]{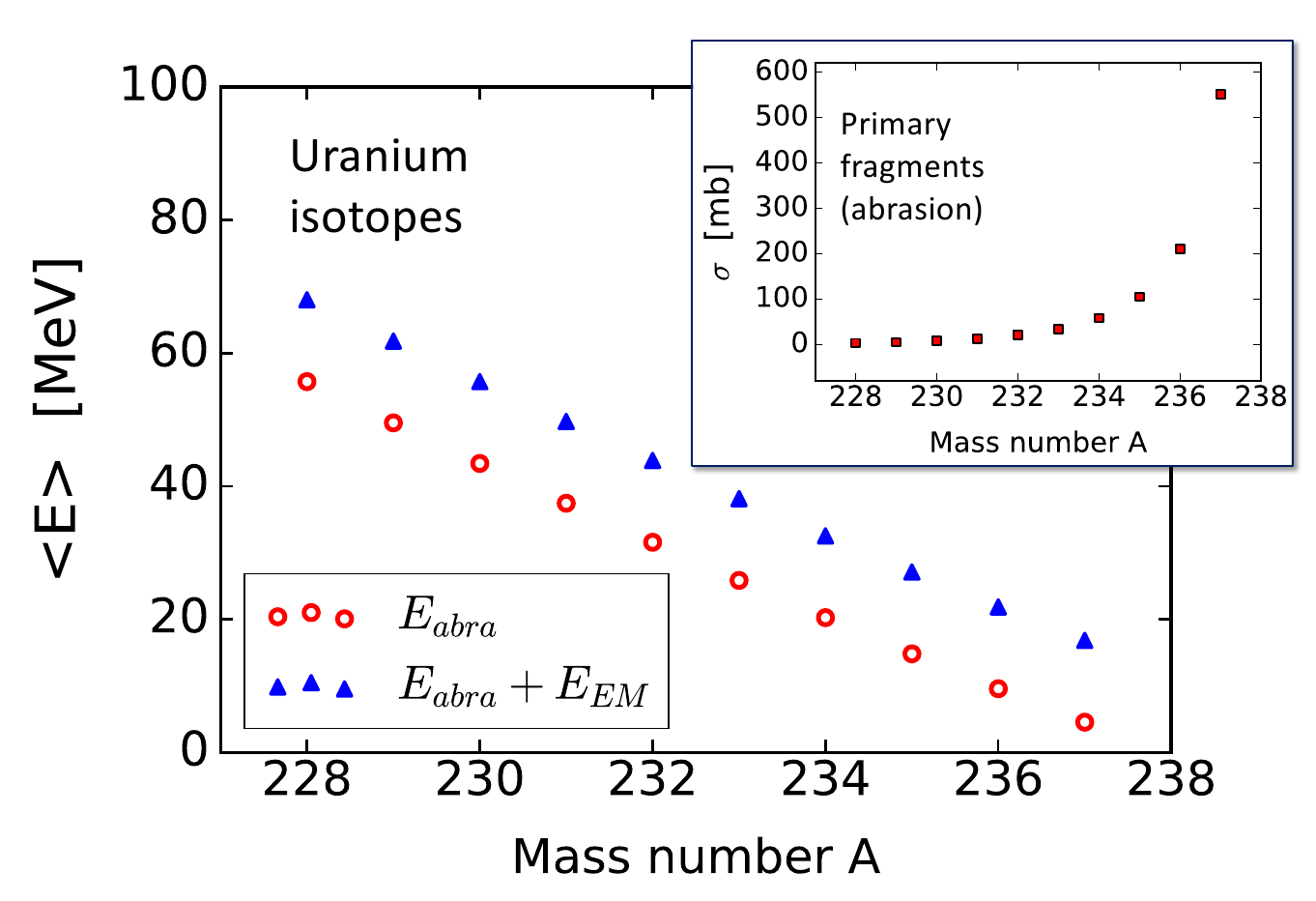}
\caption{Average excitation energies of a few uranium fragments due to abrasion (open circles) and when the average electromagnetic excitation energy of the outgoing fragments is added to them (triangles).  {\it Inset:} Cross sections for production of  primary uranium fragments  due to abrasion. \label{excenerg}}
\end{center}
\end{figure} 

The excitation of giant resonances (GR) by the nuclear and Coulomb interactions  will also lead to large cross sections. Recent progress in microscopic calculations has been made using time-dependent density-functional theories leading to fission an other decay channels \cite{StetcuPRC.2011,SimenelPRC.89.031601,BulgacPRL.116.122504,StetcuPRL2015,TanimuraPRL.118.152501}. Here we adopt a simpler picture and assume an IVGDR located at $E_{IVGDR} = 31.2A^{-1/3} + 20.6A^{-1/6}$ MeV and the DGDR located at twice the IVGDR energy  \cite{BERTULANI1988299,Aumann1995,BERTULANI1999139}. The ISGQR and IVGQR states are located at $E_{ISGQR}=62A^{-1/3}$ MeV and $E_{IVGQR}=130A^{-1/3}$ MeV, respectively. The resonances are assumed to  fully  exhaust the isoscalar and isovector  operator sum-rules \cite{SATCHLER1987215,BERTULANI1999139} and the DGDR is assumed to have the same strength for the E1 excitation GDR $\rightarrow$ DGDR.  The optical potential in Eq.  \eqref{VfiN} is generated from the t$\rho\rho$ approximation  \cite{BERTULANI1999139}  and a deformation parameter $\delta_{2}=0.438$ fm was used. The excitation probabilities from the solution of the coupled equations \eqref{rcc} yield cross sections equal to 408.2 mb for the ISGQR, 531.0 mb for the IVGQR, 4183 mb for the IVGDR and 227.3 mb for the DGDR.  The nuclear excitation contribution to the ISGQR is 18.72 mb (using Eq. \ref{VfiN}), much smaller than those stemming from EM excitation. The abrasion, EM and nuclear excitation processes are added separately and altogether they produce primary fragments with a total cross section of 13.41 b. 

{\it Ablation with EM post-excitation.} 
In the abrasion stage, each fragment acquires an excitation energy  obtained by subtracting the single particle energies for each fragment  from those of the $^{238}$U projectile. The vacant single particle (s.p.) energies are weighted with the corresponding nucleon removal probabilities for a given impact parameter.  Without meaningful loss of accuracy, we use the deformation parameter $\beta=0$ to generate the s.p. states of all fragments. For collective EM and nuclear excitation,  the energy deposited in the projectile is  obtained by multiplying the energies of each resonance $\alpha$ with the probabilities  $|{\cal A}_{\alpha}(z=\infty,b)|^{2}$ from Eq. \eqref{rcc}. The excitation energies of the primary fragments  are  inputs for the ablation stage  \cite{PhysRev.57.472}. 

\begin{figure}
\begin{center}
\includegraphics[scale=0.35]{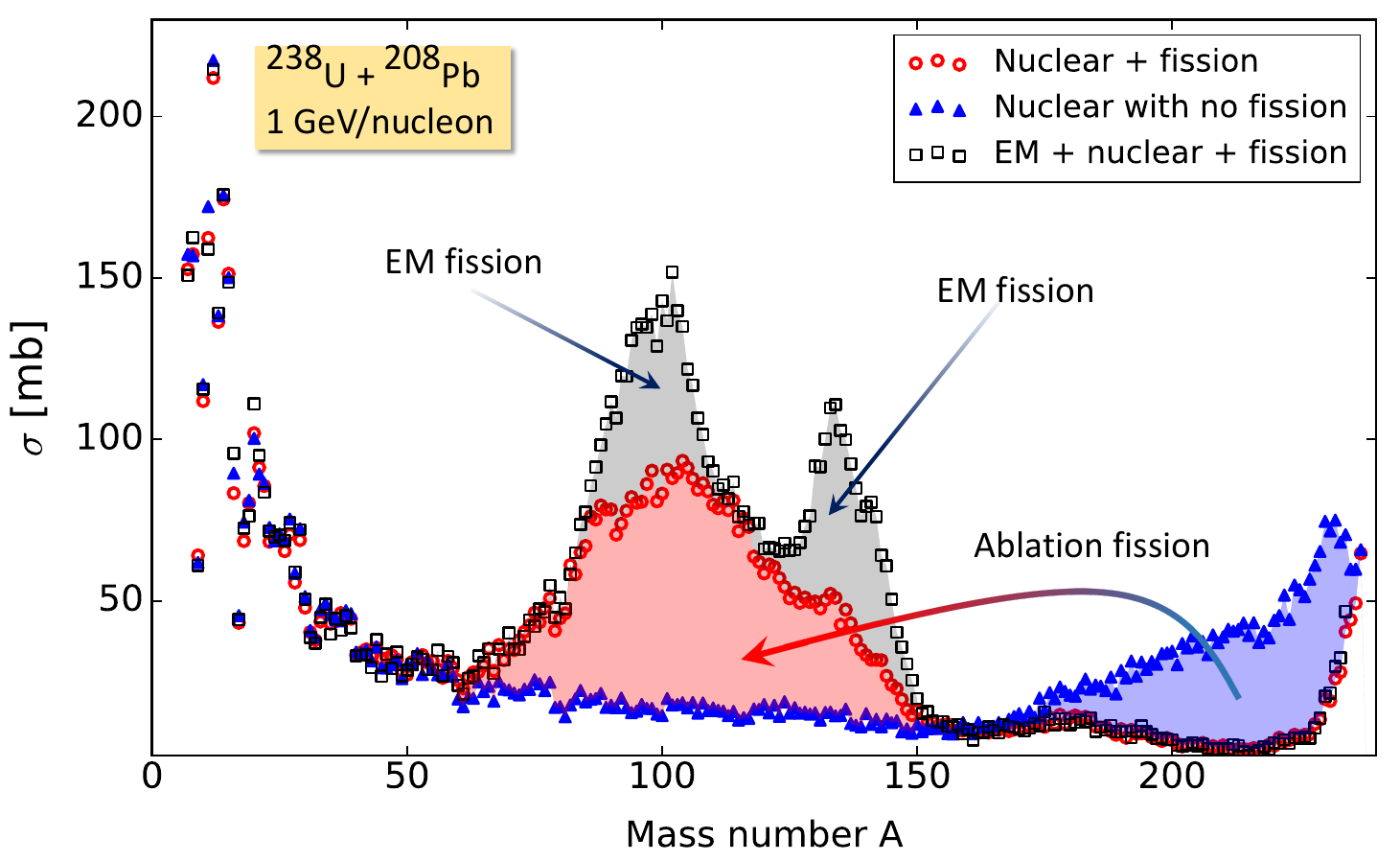}
\caption{Mass distribution of projectile fragments. The blue diamonds correspond to the yields obtained with the abrasion-ablation model without fission, while the red circles include fission decays. The black squares include the electromagnetic excitation leading to particle evaporation and fission products. \label{upbfis}}
\end{center}
\end{figure} 

We introduce a  new mechanism  to account for the excitation energy of the fragments.  It has been known for a long time that in the fragmentation of relativistic nuclei, such as $^{132}$Sn \cite{PEREZLOUREIRO2011552}, $^{136}$Xe \cite{BenlliurePRC.78.054605}, $^{197}$Au \cite{BENLLIURE199987,SCHMIDT1992699}, $^{208}$Pb \cite{KurtukianPRC.89.024616}, and $^{238}$U \cite{BENLLIURE1998458},  the energy deposited in the nucleus is not enough to explain the fragment yields.  This is an old problem discovered in early calculations where agreement with experimental data was surprisingly good \cite{GAIMARD1991709,Lanzano1992} if the excitation energy in the original abrasion-ablation model was multiplied by a factor two to three. We show that this multiplication factor is not necessary at least for reactions involving heavy nuclear targets. The idea simply relates to the large EM excitation  probabilities, close to unity, for the abraded fragments on their way out from the strong interaction region.

\begin{figure}
\begin{center}
\includegraphics[scale=0.45]{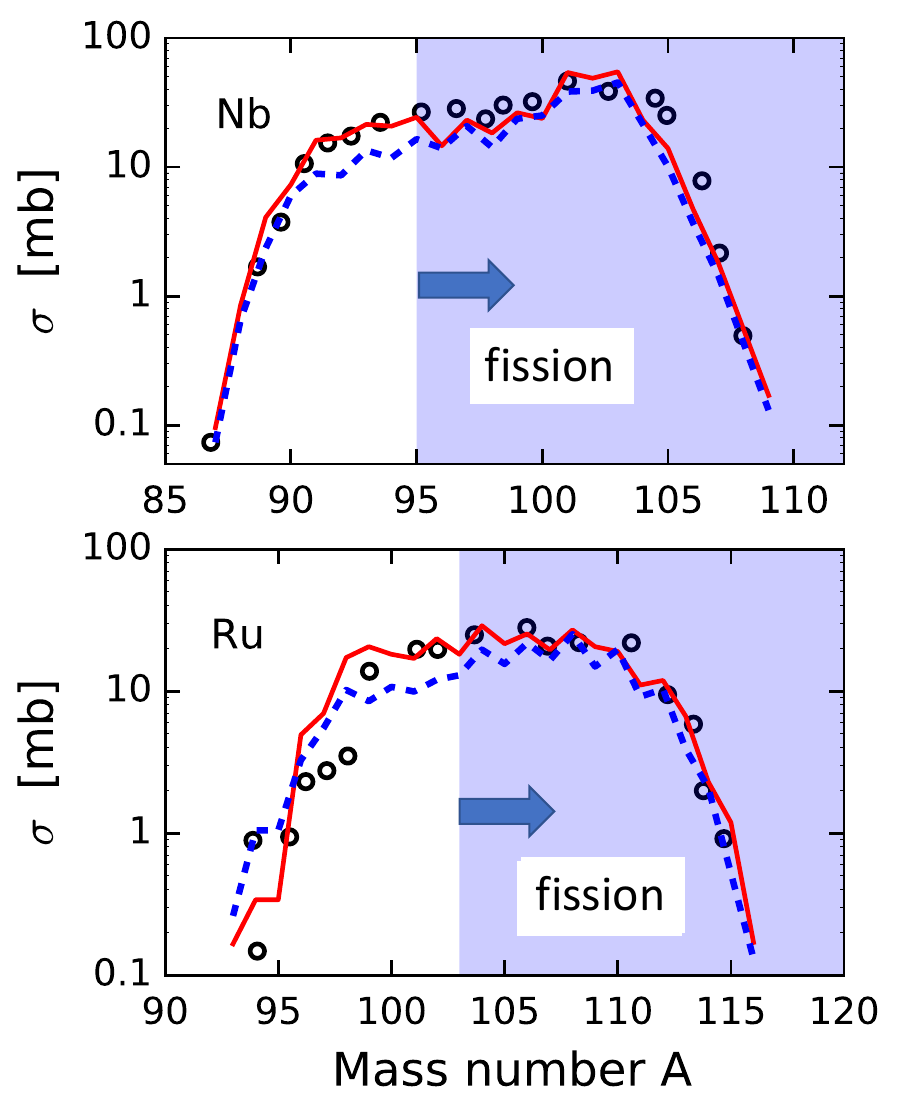}
\caption{Isotopic distribution of niobium (upper panel) and ruthenium (lower panel) fragments. Data (open circles) are from Ref. \cite{ENQVIST199947}. The solid (dashed) lines correspond to calculations with (without) inclusion of final state electromagnetic excitation of abraded fragments. The arrows point to the region (shaded area) of increasing contribution of  fragments decaying by fission. \label{upbfis2}}
\end{center}
\end{figure} 

Using Eq. \eqref{rcc} we have calculated the probabilities of EM excitation of fragments on the outgoing part of their trajectories and weighted with the giant resonance energies, adding the result to the fragment abrasion energy. Figure \ref{excenerg} shows the average excitation energies of a few uranium fragments due to abrasion (open circles) and when the average electromagnetic excitation energy of the outgoing fragments is included (triangles).  The additional EM energy is nearly constant ($11.1 - 12.5$ MeV) for the uranium fragments displayed. This effect will have the largest impact on the heaviest uranium isotopes because their production cross sections  are largest, as displayed in the smaller frame in Fig.  \ref{excenerg}. Similar results were found for protactinium, thorium, actinium, radium, and other heavy element isotopic chains. This has a visible impact on their decay to secondary fragments during the ablation stage.

The cross sections for mass distribution of each fragment calculated according to the evaporation model after ablasion and EM excitation are displayed in Fig. \ref{upbfis}. The blue diamonds (red circles) show the results without (with) inclusion of fission channels. The black squares include large impact parameters with only EM excitation. It is clear that fission products after the abrasion originate from primary fragments with mass $A\gtrsim 170$. Not shown in the figure are cross sections for primary fragments with mass $A=235 - 237$ of about $196-726$ mb, contributing mostly to fission fragments. About 23\% of the primary fragments yield fission products after ablation. These are seen as a clear bump in the plot, peaked around mass $A=110$. On the other hand, the fission yields originating from EM excitation of the $^{238}$U projectile is responsible for the double hump structure in the figure, characteristic of fission, peaked around masses $A=100$ and $A=140$.  Fission products correspond to about 18\% of the EM excitation  of $^{238}$U at large impact parameters.

Figure \ref{upbfis2} displays  the isotopic distribution of niobium (upper panel) and ruthenium (lower panel) fragments. Data (open circles) are from Ref. \cite{ENQVIST199947}. The solid (dashed) lines correspond to fragmentation calculations with (without) inclusion of final state EM excitation of abraded fragments. The arrows point to the region (shaded area) of increasing contribution of fragments decaying by fission. The overall agreement with the experimental data is reasonable and differences in particular cases are estimated to be within a factor of two to ten.  But other clear trends, also obtained in our calculations for other isotopic chains, are that: (a) Evidently, the addition of final state EM excitation of primary fragments leads to an enhancement of the average yields across the isotopic chains. (b) The yields of lighter fragments become larger when the final state EM excitations are accounted for.  This post-excitation effect, so far neglected in previous works is probably not going to provide an accurate description of the experimental data, but is clearly important  to extract useful information about fission dynamics in relativistic heavy ion collisions.

The post-excitation of primary fragments could become a useful tool to determine low limits for fission times by observing their impact on isotopic distribution of secondary fragments. Typical EM collision times  during post-excitation of fragments are $\Delta t \sim 100$ fm/$c \sim 10^{-21} $ s. A few excited primary abrasion fragments are expected to decay by fission before they can be excited again thus having an impact on their fission yields. For example, fission rates of $^{235}$U are expected to have transient and saddle-to-scission times also of the order of $10^{-21}$ \cite{JACQUET2009155}.  The changes in fission yields could be small but quantifiable as the experimental  techniques and theoretical models become more advanced.

\begin{figure}
\begin{center}
\includegraphics[scale=0.35]{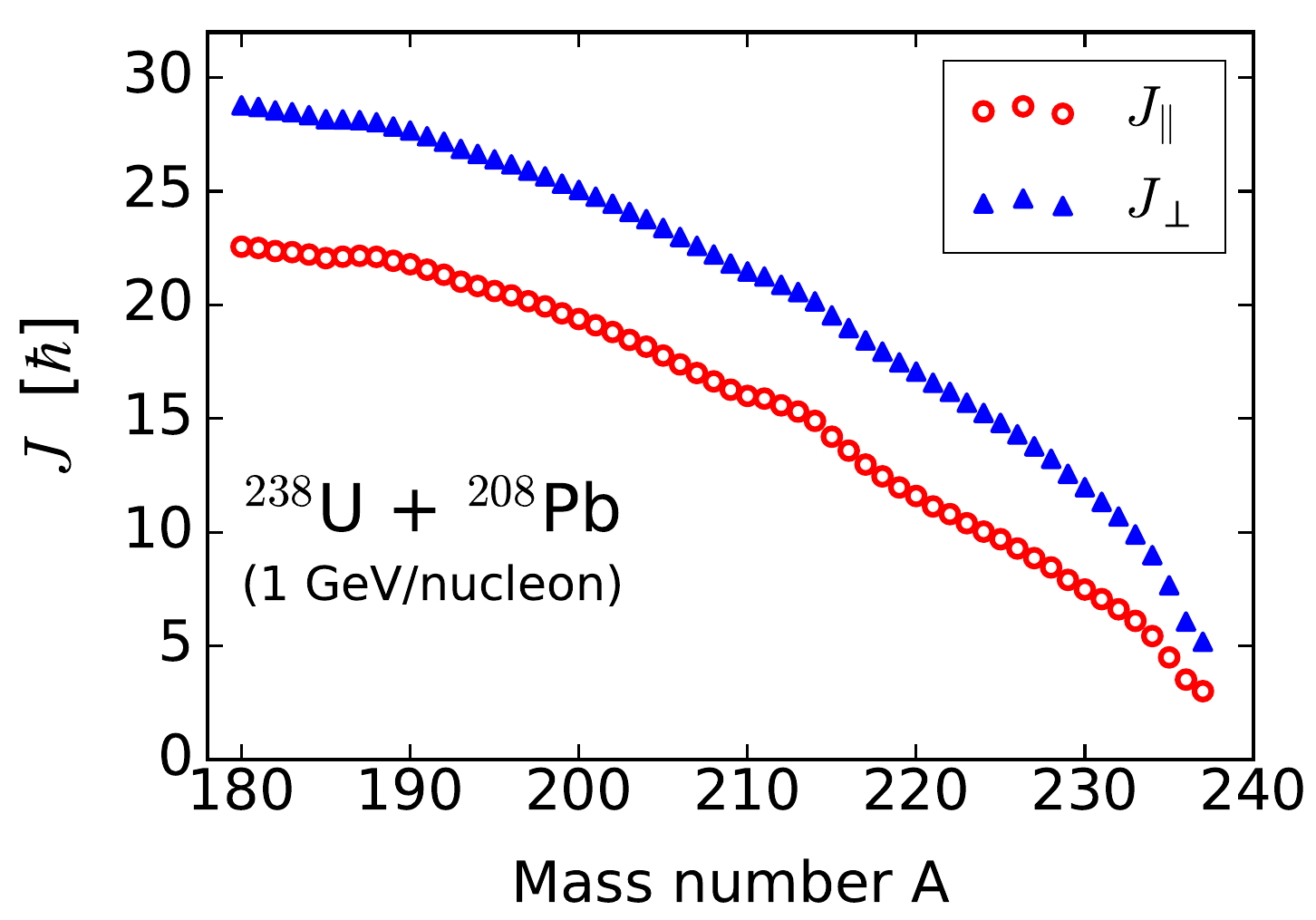}
\caption{Transverse, $J_{\perp}$, and longitudinal, $J_{\|}$, angular momentum transfer  (see text for definitions)  to primary fragments due to nucleon abrasion as a function of the fragment mass number. \label{angmom}}
\end{center}
\end{figure} 

{\it   Fragment reorientation.} Large cross sections are also obtained for EM excitation of rotational states in deformed projectiles and of abrasion products. Although the energy transfer to rotational states, a few 100 keV, has little influence on the average EM excitation energy of GRs and in the abrasion products, they could affect considerably the angular momentum transfer to the primary fragments. Using Eq.  \eqref{rcc} for a symmetric rigid rotor, one can write the EM operator as 
\begin{equation}
{\cal M}_{E2m}={e\over 2}Q_{2}Y_{2m}(\theta,\phi),
\end{equation} 
where $Q_{2}$, the intrinsic quadrupole moment, is related to the reduced matrix elements by 
\begin{equation}
Q_{2}^{2}={16\pi \over 5c^{2}} B(E2;0^{+}\rightarrow 2^{+}).
\end{equation} 
For the quadrupole moment of $^{238}$U we use $Q_{2}=11\times 10^{3}$ fm$^{2}$  \cite{RAMAN19871}. We consider the excitation of states in the ground state rotational band, starting from the rotation angular momentum  $I= 0$ of the ground state. We obtain cross sections of 3.21 b for excitation of $I=2$ and only 41.7 mb for the $I=4$ rotational states. Since the probability for EM excitation of GRs is nearly one at grazing collisions, a large cross section also emerges for the excitation of rotational states on top of the IVGDR, ISGQR and IVGQR. Total cross sections of 1.24 b for the $\big| IVGDR\big> \bigotimes \big| I=2\big>_{{rot}} $  and 35.5 mb for $\big| IVGDR\big> \bigotimes \big| I=4\big>_{{rot}} $ were obtained, which are the largest cross sections for rotational states built on GRs.  Thus, at least three extra units of angular momentum, $J =3\hbar$, are expected to build up on the projectile before its decay. This angular momentum transfer has a small \footnote{Sizable effects on the effect potential energies are expected for $J\sim 30 \hbar$.}, but non-negligible consequence for the effective potential energies and their dependence on the neck formation radius along the fission path. The same conclusion applies to EM reorientation of primary fragments after the abrasion process.

Because of the sudden removal of nucleons in close collisions, angular momentum is also transferred to the fragments during abrasion. A simple estimate can be done by using the model described in Ref. \cite{Abul-Magd1976} for the average momentum transfer, $\Delta p$, together with the Newtonian estimate $J=\Delta p R$  for the average angular momentum transfer, where $R$ is the projectile radius. Our results are shown in Fig. \ref{angmom}, were $J_{\|}$ ($J_{\perp}$) is the angular momentum transfer associated to the longitudinal (transverse) linear momentum transfer to the projectile when nucleons are abraded. In collisions at high energies, the transverse momentum transfer is always larger than the longitudinal one. The total average angular momentum transfer can be inferred from $J=\sqrt{J_{\|}^{2}+J_{\perp}^{2}}$.  Only a moderate amount ($J\sim 3-5\hbar$) of angular momentum  is transferred to large mass fragments due to abrasion, such as $^{237}$U, but as the fragment mass decreases the angular momentum deposited will lead to a large modification of the effective potential energies at the scission point and the ensuing fission yields. This effect has been overseen in previous works. But it also means that a better description of the decay of the primary fragments needs to be done, beyond the simplified Ewing-Weisskopf approach which neglects angular momentum effects in the decay process. The Newtonian estimate is not reasonable when many nucleons are abraded and we have avoided to show results for light fragments, which probably can be better described using  intranuclear cascade (INC) models \cite{MancusiPRC.91.034602,RS.PRC.96.054602}. INC models can also handle pion production and propagation, which changes the excitation energy.

{\it Summary and conclusions.}  We propose a new method to compute fission and fragmentation cross sections in relativistic heavy ion collisions. Our method links the isotopic distribution of fragments with the physics  of nucleon removal in the binary nucleon-nucleon collisions and electromagnetic excitation of giant resonances.  We add to this scenario the hitherto neglected physics of fragment excitation and reorientation in the final state. We believe we have uncovered these  unappreciated reaction mechanisms previously buried in the standard two step mechanism of primary fragment production followed by particle evaporation and fission. The reaction mechanism we propose is independent of other effects already accounted for. It may be needed to reach the fragment excitation energies required to explain known differences between theory and experiment, at least for heavy nuclear targets. We demonstrate that it also plays a role in the build up of angular momentum of fragments prior to their decay to be studied in future experiments \footnote{Radomira Lozeva et al., work in progress}.

The design of new detectors and increase of relativistic radioactive beam intensities provides a unique opportunity to study the important physics of fission of numerous radioactive nuclides,  beyond the reach of fixed target experiments. Undoubtedly, previous studies have shown the usefulness of the method and the complementary role it plays in understanding the dynamics of fission \cite{Andreyev_2017,Schmidt_2018,SCHMIDT2000221,SCHMIDT1994313,Aumann1995,Armbruster1996,BENLLIURE1998458,SCHMIDT2000221}. We noticed that improvements can be done in presently adopted theories without which an accurate description of the experimental data at the level required for the study of fission dynamics is not possible.

{\it Acknowledgement.}  We have benefited from useful discussions with Aleksandra Kelic (GSI)  and Jose Benlliure (Santiago de Compostela) on the abrasion-ablation model. This work has been supported by the Turkish Council of Higher Education (YOK) under Mevlana Exchange Project Number MEV-2019-1744, and by the U.S. DOE grants DE-FG02-08ER41533 and  the U.S. NSF Grant No. 1415656.


\end{document}